\documentstyle[11pt,Cargesepasp,twoside,epsf]{article} \def\edcomment#1{\iffalse\marginpar{\raggedright\sl#1\/}\else\relax\fi}
\reversemarginpar

\begin{document}
\title{Statistical properties of visual binaries as tracers of the
  formation and early evolution of young stellar clusters}

\author{Gaspard Duch\^ene \& J\'er\^ome Bouvier}

\affil{Laboratoire d'Astrophysique, Observatoire de Grenoble,
  Universit\'e Joseph Fourier, B.P. 53, 38041 Grenoble Cedex 9,
  France}

\author{Jochen Eisl\"offel}

\affil{Th\"uringer Landessternwarte Tautenburg, Sternwarte 5, 07778
Tautenburg, Germany}

\author{Theodore Simon}

\affil{Institute for Astronomy, University of Hawaii, 2680 Woodlawn
Drive, Honolulu, HI 96822, USA}
 
\begin{abstract}
  In this review, we summarize the observed statistical properties of
  visual binaries in various young stellar clusters. The binary
  frequency, orbital period distribution and mass ratio distribution
  are considered for populations of both low-mass and high-mass
  stars. These properties are then compared to numerical models of the
  dynamical evolution of stellar clusters and to the predictions of
  binary formation mechanisms.
  
\end{abstract}

\section{Introduction}

The current paradigm for star formation is that most stars form in
stellar clusters rather than in Taurus-like star-forming regions.
Since binary systems are more frequent than single stars, the study of
multiple systems in clusters is expected to provide us with some
insight on the stellar formation process itself. For instance, it has
been suggested that the gravitational collapse of molecular clouds
leads to a binary fraction close to 100\,\%, which would decrease over
time to about 60\,\%, the value observed on the Main Sequence (Ghez
{\it et al.} 1993). The physical processes leading to such an
evolution and the timescale over which they act still need to be
asserted. Alternatively, the impact of local physical conditions in
the molecular cloud, such as the magnetic field strength or gas
temperature, on the output of the star formation process has still to
be ascertain and might lead to different binary frequencies in various
environments (Durisen \& Sterzik 1994).

Observing stellar clusters with a wide range of physical parameters
allows to perform a direct comparison between binary properties and
model predictions. Clusters with ages ranging from one to several
hundred million years are known within 500\,pc from the Sun with some
hosting many massive stars while other containing only low-mass stars.
The stellar density can vary over more than an order of magnitude from
cluster to cluster, as well as the physical conditions of the parent
molecular cloud. In this contribution, we review the observed
properties of visual binaries in clusters and compare them with
cluster properties and with model predictions regarding the formation
and evolution of stellar clusters.

\section{Summary of observational results}

\subsection{Visual binary frequency}

The advent of several high-angular imaging techniques makes the binary
frequency a relatively straightforward property to derive in stellar
populations.  For instance, $HST$ imaging (Prosser {\it et al.} 1994),
speckle interferometry (Leinert {\it et al.} 1993, Patience {\it et
  al.} 1998) and adaptive optics (Bouvier {\it et al.}  1997) have
been used, mostly in the near-infrared in order to facilitate the
detection of relatively faint and cold low-mass companions. The
low-mass populations of extremely young clusters (the Trapezium
cluster, IC\,348) as well as of more evolved ones ($\alpha$\,Per, the
Pleiades, the Hyades, Pr{\ae}sepe) have been studied, which span an
age range from $\sim$1 to several 100~Myr. The range of semi-major
axes probed by these surveys usually extends from a few 10 to
$\sim$2000 AU. The Hyades cluster, which is the closest to the Sun, is
the only one for which visual binaries could be resolved down to about
5\,AU.

Most surveys miss the faintest companions so that a statistical
completeness correction has to be applied in order to estimate the
true binary frequency from the observed one. This correction can be
large, especially at small separations for adaptive optics surveys due
to the separation-dependent detection limit inherent to this
technique. The correction factor also depends upon assumptions
regarding the distribution of mass ratios and/or orbital periods of
the binary population, which obviously increases the statistical
uncertainties on the derived binary frequency.

With these limitations, all studied clusters are found to exhibit a
visual binary frequency consistent with that of field G-dwarfs
(Duquennoy \& Mayor 1991), to within 10\,\% or less (e.g. Prosser {\it
  et al.} 1994, Bouvier {\it et al.}  1997, Duch\^ene {\it et al.}
1999). This result does not seem to depend upon the age of the cluster
between $\sim$1~Myr and 600~Myr. One possible exception might be the
Hyades cluster which appears to host more tight (5--50\,AU) binaries
than the field population does, though statistical uncertainties could
account for this slight excess (Patience {\it et al.}  1998). It hence
appears that the fraction of binaries in very young clusters, at an
age of about a million years, is already similar to that measured for
field dwarfs and that it does not significantly evolve as the clusters
dynamically relax on the ZAMS.

\subsection{Orbital period distribution}

While the overall binary frequency integrated over a range of
semi-major axes already provides clues to the properties of binary
populations in clusters, one would ideally wish to compare the
distribution of their orbital periods. One is thus led to convert the
observed angular separations into orbital periods assuming an average
system mass and applying a statistical correction factor to take into
account projection effects.

\begin{figure}[t]
\plotfiddle{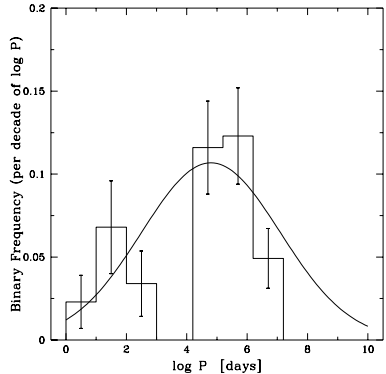}{4.5truecm}{0}{26}{26}{-70}{-40}
\caption{The distribution of orbital periods for Pleiades G and K-type 
  binaries. Spectroscopic binaries are taken from Mermilliod {\it et al.}
  (1992) and visual binaries from Bouvier {\it et al.} (1997). The solid
  curve is the gaussian fit to the field G dwarf binary distribution
  (Duquennoy \& Mayor 1991). Both the shape of the distribution and the
  fraction of binaries in each orbital period bin appear similar in
  Pleiades and field binaries. }
\end{figure}

Over a limited range of orbital periods, the distribution of
semi-major axes in cluster binaries again appears to be consistent
with that of field G-dwarf binaries (Duquennoy \& Mayor, 1991). The
comparison between cluster and field binaries can be further
investigated by considering not only visual binaries but spectroscopic
binaries as well. This has been done for ZAMS clusters, where the
low-mass stars have already been searched for short-period companions.
Within statistical uncertainties, the results reinforce the similarity
between cluster and field binaries, as shown for the Pleiades in
Figure\,1.

\begin{figure}[t]
  \plotfiddle{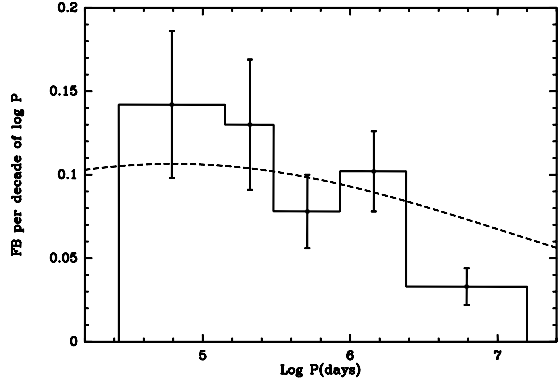}{4.5truecm}{-90}{28}{28}{-105}{150}
\caption{The orbital period distribution for solar-type binary systems
  in three main sequence clusters ($\alpha$\,Per, the Pleiades and
  Pr{\ae}sepe). To ensure identical detections limits for all
  clusters, only results from adaptive optics surveys are considered
  here. Completeness corrections mostly affect the first period bin,
  and are similar for all clusters.  Error bars represent 1\,$\sigma$
  Poisson uncertainties. The dashed curve is the gaussian fit to the
  field G-dwarf binary population (Duquennoy \& Mayor 1991).}
\end{figure}

In order to improve the statistics, we combined the results obtained
for several ZAMS clusters. This yields the orbital period distribution
shown in Figure\,2 for visual binaries. While the distribution of
period in cluster binaries agrees with that of field binaries over the
orbital range $10^{4.4}$--$10^{6.4}\,$days, a paucity of wide (sep
$\geq$ 300 AU) cluster binaries appears relative to the field at the
2.5\,$\sigma$ level. Such a deficit cannot be related to a detection
bias, since the distant companions of wide binaries are the most
easily detected. The uncompleteness correction which was applied here
affects only the first orbital bin of the distribution.

This difference between field and cluster binaries is reinforced by
the evidence that the orbital period distribution peaks at a smaller
value in clusters than in the field (Patience {\it et al.}  2000).
The mode of the distribution is 5\,AU for cluster binaries and 40\,AU
for field binaries, the former being estimated under the assumption
that all clusters, including the Hyades, can be combined into a single
sample. Furthermore, Scally {\it et al.}  (1999) have shown that there
are no very wide binaries (sep $\geq$ 2000 AU) in the Trapezium
cluster.

Hence, even though clusters at various evolutionary stages appear to
host a roughly similar fraction of low-mass binary systems, which
itself is similar to the field binary fraction, the distribution of
orbital periods appears to exhibit small but significant differences
between clusters and field binary populations.

\subsection{Mass ratio distribution}

The mass of the primaries and secondaries of binary systems on the
ZAMS can be reliably estimated from their near-infrared flux using
main sequence mass-luminosity relationships. In order to increase the
statistical significance of the derived mass ratio distribution, the
combined results for three clusters observed with similar set-ups are
shown in Figure\,3.  The observed distribution appears to be
essentially flat over the range $q=M_b/M_a=0.1$--1. This agrees with
the $q$-distribution derived by Duquennoy \& Mayor (1991) for low-mass
field binaries if the limited sensitivity of our observations, which
prevents us from detecting tight systems with low mass ratios, is
taken into account.

\begin{figure}[t]
\plotfiddle{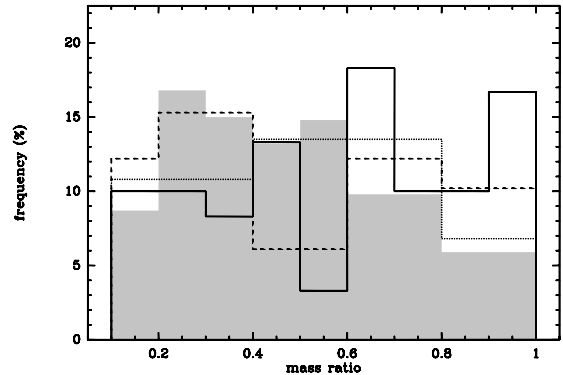}{4.5truecm}{-90}{28}{28}{-105}{150}
\caption{The mass ratio distribution derived for various populations
  of low-mass binaries. {\it Shaded histogram}: field G-dwarf
  distribution from Duquennoy \& Mayor (1991) corrected for the
  detection limits of current adaptive optics binary surveys in ZAMS
  clusters. This correction removes some of the low-$q$ binaries, and
  thus flattens the intrinsic distribution. {\it Solid histogram}:
  adaptive optics observations of ZAMS cluster members ($\alpha$\,Per,
  Pleiades, and Pr{\ae}sepe), without applying any completeness
  correction; {\it Dashed histogram}: T\,Tauri stars in Taurus from
  Leinert {\it et al.}  (1993); {\it Dotted histogram}: low-mass
  members of the Trapezium cluster from Prosser {\it et al.}  (1994).
  All distributions are undistinguishable within statistical error
  bars.}
\end{figure}

Contrary to main sequence binaries, young low-mass systems frequently
exhibit significant continuum excesses at near-infrared wavelengths
and this may affect the derivation of mass ratios from near-IR flux
ratios.  However, it is usually assumed on a statistical basis that
both stars in a young binary system exhibit similar excesses, and that
the overall flux ratio distribution can be safely transformed into a
mass ratio distribution.  Baraffe {\it et al.} (1998)'s models were
used to get the mass-luminosity relationship relevant for such young
stars. Both in the Taurus star-forming region (Leinert {\it et al.}
1993) and in the Trapezium cluster (Prosser {\it et al.} 1994), the
observed $q$-distribution is consistent with being flat down to
$q=0.1$, and therefore not different from the ZAMS and field
estimates.

Current results thus suggest that the mass ratio distribution is
similar for all low-mass binary populations studied so far,
independent of their evolutionary status and environmental conditions
(field or clusters).  No peak is found in the distribution towards
$q\approx1$ nor towards low $q$.  Furthermore, the ZAMS clusters mass
ratio distribution does not seem to vary with separation for visual
binaries (Patience {\it et al.} 2000).

\subsection{Massive binaries}

Studies conducted so far have mostly focussed on the low-mass binary
population of clusters. More recently, however, the investigation of
massive binaries has started, with the additional difficulty of a huge
contrast between the bright primary and the low-mass companion which
is a major limitation of current massive binary surveys.

The recent {\it HST} study of thirteen OB stars in the Trapezium
cluster by Preibisch {\it et al.} (1999) already revealed eight visual
companions with separations ranging from 15 to 450\,AU. The binary
frequency of high-mass stars in this cluster thus appears
significantly higher than that of their low-mass counterparts (Prosser
{\it et al.} 1994), and the difference is further enhanced when
completeness corrections are taken into account.  Similarly, an
adaptive optics survey of about fifty OB stars in the distant cluster
NGC\,6611 revealed eight companions in the separation range
200--2000\,AU (Duch\^ene 2000), indicative of a high binary fraction
compared to G-type field dwarfs if the orbital period distribution is
the same for high-mass and low-mass binaries. A study of young, mostly
isolated, Herbig AeBe stars conducted by Bouvier \& Corporon (2000)
yields an uncorrected binary frequency which is significantly higher
than that of lower mass Main Sequence stars for both Be and Ae PMS
stars.

\begin{figure}[t]
  \plotfiddle{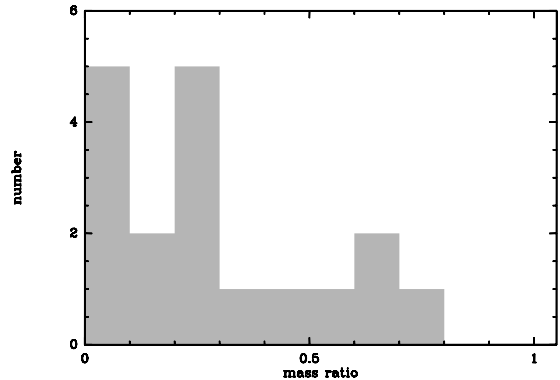}{4.5truecm}{-90}{28}{28}{-105}{150}
\caption{Mass ratio distributions for young B stars located in the
  Trapezium cluster (Preibisch {\it et al.} 1999), in NGC\,6611
  (Duch\^ene 2000) and isolated Herbig Be stars (Bouvier \& Corporon
  2000). Low $q$ are much more common in high-mass binaries than in
  solar-type binaries.  However, there are hints that the mass ratio
  distribution moight be different for cluster and isolated young
  massive stars.}
\end{figure}

The mass-ratio distribution of massive binaries can be estimated from
the flux ratios measured by imaging surveys. The $q$-distribution
derived for massive binaries with B0--B7 primaries is shown in
Figure\,4. The sample includes members of the Trapezium cluster and
NGC 6611 as well as isolated Herbig Be stars, with a semi-major axis
ranging from 100 to 2000\,AU.  The mass ratio distribution is strongly
peaked towards small values: only about 25\,\% of the binaries have
$q>0.5$. Patience {\it et al.} (2000) obtained a similar trend for A
and F stars in various ZAMS clusters. This distribution is
dramatically different from that of low-mass binaries, indicating that
the primary's mass is a relevant parameter when investigating the
formation and evolution of binary systems.

\section{Interpretation}

The observed properties of young binary systems hold clues to their
formation mechanism and their early evolution. In the following, we
consider both classes of models, binary formation and dynamical
evolution of clusters, and compare their predictions with
observations.

\subsection{Dynamical evolution of stellar clusters}

The fact that all clusters studied so far exhibit a similar fraction
of visual binaries regardless of their age rules out a long-term
decrease of the binary frequency past the first million years after
gravitational collapse. In addition, since similarly-aged loose
associations and clusters display different binary frequencies, it is
tempting to conclude that environmental effects play a key role in
fixing the binary content of a star forming region.

$N$-body simulations of the dynamical evolution of stellar clusters
were ran by Kroupa {\it et al.} (1999) and compared with observations
of the Trapezium cluster. They found that close encounters between
cluster members could efficiently disrupt binaries in less than a
million years while the number of binaries would remain constant
afterwards. This results from the fact that the number of encounters
falls off dramatically after one relaxation time. If the initial
low-mass binary fraction is of the order of 100\,\% at the end of the
protostellar collapse, such a scenario would naturally explain the
high binary frequency observed in T\,Tauri associations: in these
regions, the stellar density is several orders of magnitudes too low
to generate numerous enough close-by interactions so as to impact on
the initial binary content (Clarke \& Pringle 1991). In contrast,
frequent gravitational encounters in dense clusters would rapidly
erode the initially high binary fraction and lead to a binary
frequency similar to that of the field after a million years of
evolution or so.

\begin{figure}[t]
  \plotfiddle{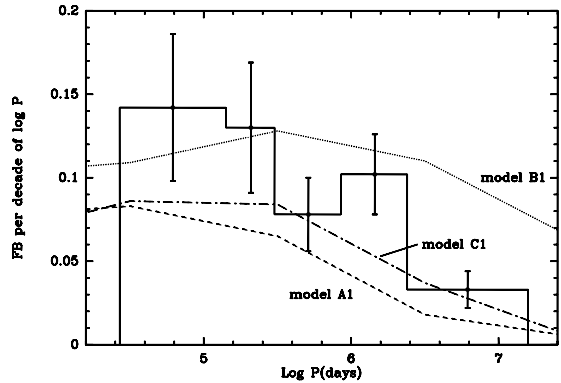}{4.5truecm}{-90}{28}{28}{-105}{150}
\caption{Comparison of the observed orbital period distribution for
low-mass stars in ZAMS clusters to the end state of Kroupa {\it et
al.} (1999)'s simulations of the dynamical evolution of a
Trapezium-like cluster.  Models represent respectively a virial
equilibrium (model A1), a rapidly expanding (model B1) and a
collapsing cluster (model C1). Most simulations disrupt wide binaries,
in good agreement with the observations, to the exception of model B1
which only disrupts those binaries with orbital period larger than
about $10^{7.5}$ days.}
\end{figure}

The observed properties of binary populations described in Section\,2
further support the hypothesis that dynamical evolution of clusters
may largely drive the binary properties. In dynamical models, wide
binaries are preferentially disrupted, because their binding energy is
small compared to the average kinetic energy of the cluster (see the
review by P. Kroupa in this volume). This prediction agrees with the
observed distribution of orbital periods for binaries in ZAMS clusters
which exhibits a paucity of wide systems, as shown in Figure\,5. The
higher binary fraction found for massive stars is also in qualitative
agreement with this model since the potential well of massive
primaries is deeper than that of solar-mass stars, making it more
difficult to disrupt more massive binaries. The ``softening boundary''
which delineates hard and soft systems, the latter being prone to
disruption, is thus moved towards wider separations.

A number of issues remain however to be addressed before accepting
this paradigm. For instance, dynamical models predict that low-$q$
systems are preferentially disrupted, introducing a non-negligible
modification in the shape of this distribution. This constrasts with
the derivation of similar mass ratio distributions for low-mass
binaries in clusters and associations. Quantitatively, Kroupa (1995)
estimated that 70\,\% of the initial binaries must have a mass ratio
$q<0.4$ in order to account for the observed distribution of field
binaries. This is unlike the $q$-distribution observed for T\,Tauri
associations. Another puzzling result is the fact that all clusters
seem to have very similar binary frequency and orbital period
distributions. Hence, binary properties appear to be weakly dependent
on small differences in initial conditions. Dyamical models have
focussed on reproducing the binary population of the Trapezium cluster
and it remains to be seen whether the same results would be obtained
for slightly less dense clusters such as IC\,348.

According to these models, wide binaries found in the solar
neighbourhood cannot have formed in clusters, since they would have
been rapidly disrupted. One thus has to assume that they were born in
loose associations that have subsequently dissipated in the field.
However, the field binary population does not appear to be a simple
combination of cluster and association members: a weighted mixture of
these 2 populations that reproduces the observed fraction of tight and
wide binaries also yields a large excess of systems in the 5--50\,AU
separation range compared to field binaries. The issue of the
efficiency of star formation in clusters as opposed to associations is
still open.

\subsection{Binary formation models}

If the dynamical evolution of clusters is responsible for the observed
binary properties, then the formation process itself must yield a
large initial binary fraction. Furthermore, the masses of the
components must be taken at random from the initial mass function, as
suggested by Kroupa {\it et al.} (1995). We note that the observations
of high-mass binaries are qualitatively consistent with random
pairing, with a limited number of $q>0.5$ pairs, though this result is
still preliminary due to the small sample size. We will now discuss
whether binary formation mechanisms do predict such properties for
initial binaries, prior to cluster's evolution.

The formation of a stellar cluster probably results from the
fragmentation of the parent molecular cloud into a large number of
individual cores.  In this environment, wide binary systems can form
either by capture or through the fragmentation of a small collapsing
core. Models considering capture in small-$N$ ($N<10$) subclusters
predict that the two most massive members are usually associated in a
binary system, while the other stars are ejected (McDonald \& Clarke
1993). This leads to a mass ratio distribution which either peaks
towards unity or is roughly flat, depending on the value of $N$.
However, taking into account the dissipative effect of protostellar
disks during close encounters leads to a secondary mass distribution
that mimics the IMF (McDonald \& Clarke 1995).  Furthermore, the model
predicts a higher binary frequency for more massive stars. Similar
simulations made by Sterzik \& Durisen (1998) additionally predict
that the mass ratio distribution is more heavily weighted towards
low-$q$ values in higher mass systems. All these predictions are
consistent with the observations, indicating that stars in clusters
may form in compact subclusters.  However, such a scenario does not
end up with a 100\,\% initial binary frequency, since some single
stars are ejected from each subcluster.

Capture during the relaxation of small-$N$ subgroups appears to be the
model that best fits current observations, and especially the random
pairing of secondary masses. The alternative model of scale-free
fragmentation (Clarke 1996) predicts that binary properties do not
depend on the primary mass, which conflicts with the observations.
Binary properties could also result, at least partly, from the
accretion of residual material onto two gravitationally bound seeds
(Bate \& Bonnell 1997, Bate 2000). However, these models predict a
larger number of $q\sim1$ binaries for tighter systems as well as for
more massive binaries, which is not verified in the observational
trends. Finally, binaries formed through disk fragmentation should
preferentially have small separations and low mass ratios, a
prediction which contrasts with the apparent lack of a dependence of
the mass ratio distribution upon separation.

\section{Summary}

The advent of high-angular resolution imaging techniques have offered
the opportunity to systematically survey binaries in young clusters,
thus complementing spectroscopic studies. We reviewed some statistical
properties of visual binaries such as their frequency, orbital period
distribution and mass ratio distribution; both low-mass and high-mass
stars were considered. Observational results can be summarized as
follows: the low-mass binaries in cluster have similar properties as
Main Sequence (field) binaries, except for a significant deficit of
systems with a separation larger than 300\,AU. High-mass stars host
many more companions than do solar-type stars, and most massive
binaries have a small mass ratio. These observational constraints are
consistent with the results of numerical simulations describing the
early dynamical evolution of stellar clusters provided that the
initial binary frequency is of the order of 100\,\% for low-mass
stars. Disk-assisted capture in small subclusters is the formation
model that best approaches the initial conditions required to account
for the observations. In both classes of models, however, several
issues still need to be investigated and additional simulations over a
wider range of physical conditions have to be run before these
scenarios can be considered as offering an acceptable new paradigm for
the star and binary formation process. In parallel, future
observations of young embedded clusters will help to derive the
properties of protobinaries at an even earlier stage of evolution than
discussed in this review, which will undoubtly lead to new clues to
their formation.


\begin{references}
  \reference Baraffe, I., Chabrier, G., Allard, F., Hauschildt, P. H.,
  1998, \aap, 337, 403

  \reference Bate, M. R. \& Bonnell, I. A. 1997, \mnras, 285, 33
  
  \reference Bate, M. R. 2000, \mnras, 314, 33
  
  \reference Bouvier, J., Rigaut, F., \& Nadeau, D. 1997, \aap, 323,
  139 

  \reference Bouvier, J. \& Corporon, P. 2000, IAU Symposium 200: The
  Formation  of Binary Stars, eds. Matthieu \& Zinnecker
 
  \reference Clarke, C. J., \& Pringle, J. E., 1991, \mnras, 249, 584

  \reference Clarke, C. J., 1996, \mnras, 283, 353
 
  \reference Duch\^ene, G., 1999, \aap, 341, 547

  \reference Duch\^ene, G., Bouvier, J., \& Simon, T. 1999, \aap, 343,
  831 
  
  \reference Duch\^ene, G. 2000, Ph.D. thesis, University of Grenoble
  
  \reference Duquennoy, A., \& Mayor, M. 1991, \aap, 248, 485
  
  \reference Durisen, R. H., \& Sterzik, M. F. 1994, \aap, 286, 84
  
  \reference Ghez, A. M., Neugebauer, G., \& Matthews, K. 1993, \aj,
  106, 2005

  \reference Kroupa, P. 1995, \mnras, 277, 1491
  
  \reference Kroupa, P., Tout, C. A., \& Gilmore, G. 1993, \mnras,
  262, 545

  \reference Kroupa, P., Petr, M. G., \& McCaughrean, M. J. 1999, New
  Astronomy, 4, 495

  \reference Leinert, C., Zinnecker, H., Weitzel, N., Christou, J.,
  Ridgway, S. T., Jameson, R., Haas, M., \& Lenzen, R. 1993, \aap,
  278, 129
  
  \reference McDonald, J. M. \& Clarke, C. J. 1993, \mnras, 262, 800

  \reference McDonald, J. M. \& Clarke, C. J. 1995, \mnras, 275, 671
  
  \reference Mermilliod, J.-C., Rosvick, J. M., Duquennoy, A., \&
  Mayor, M. 1992, \aap, 265, 513

  \reference Patience, J., Ghez, A. M., Reid, I. N., Weinberger, A.
  J., \& Matthews, K. 1998, \aj, 115, 1972

  \reference Patience, J., Ghez, A. M., \& Reid, I. N. 2000, in
  preparation

  \reference Prosser, C. F., Stauffer, J. R., Hartmann, L., Soderblom,
  D. R., Jones, B. F., Werner, M. W., \& McCaughrean, M. J.  1994,
  \apj, 421, 517

  \reference Preibisch, T., Balega, Y., Hofmann, K., Weigelt, G., \&
  Zinnecker, H., 1999, New Astronomy, 4, 351
  
  \reference Scaly, A., Clarke, C. J., \& McCaughrean, M. J. 1999,
  \mnras, 306, 253
  
  \reference Sterzik, M. F., \& Durisen, R. H. 1998, \aap, 339, 95
  
\end{references}
\end{document}